\documentclass{article}
\usepackage{spconf,amsmath,graphicx}
\usepackage{multirow}
\usepackage{makecell}
\usepackage{booktabs}

\usepackage{amssymb}
\usepackage{amsmath}

\title{LEARNING DECOUPLING FEATURES THROUGH ORTHOGONALITY REGULARIZATION}
\name{Li Wang$^1$, Rongzhi Gu$^1$, Weiji Zhuang$^2$, Peng Gao$^2$, Yujun Wang$^2$, Yuexian Zou$^{1,*}$
\thanks{
*Corresponding author \newline \indent
This paper was partially supported by Shenzhen Science \& Technology
Fundamental Research Programs (No: JSGG20191129105421211 and GXWD20201231165807007-20200814115301001)
}
}
\address{
  $^1$ ADSPLAB, School of ECE, Peking University, Shenzhen, China\\
  $^2$ Xiaomi Inc., Beijing, China
}
\begin{document}
%
\maketitle
\begin{abstract}
Keyword spotting (KWS) and speaker verification (SV) are two important tasks in speech applications.
Research shows that the state-of-art KWS and SV models are trained independently using different datasets since they expect to learn distinctive acoustic features.
However, humans can distinguish language content and the speaker identity simultaneously. 
Motivated by this, we believe it is important to explore a method that can effectively extract common features while decoupling task-specific features.
Bearing this in mind, a two-branch deep network (KWS branch and SV branch) with the same network structure is developed and a novel decoupling feature learning method is proposed to push up the performance of KWS and SV simultaneously where speaker-invariant keyword representations and keyword-invariant speaker representations are expected respectively. 
Experiments are conducted on Google Speech Commands Dataset (GSCD). The results demonstrate that the orthogonality regularization helps the network to achieve SOTA EER of 1.31\% and 1.87\% on KWS and SV, respectively.
\end{abstract}
\section{Introduction}
\label{sec:intro}
With the development of speech technology, speech assistants are expected to help people solve affairs more efficiently. People increasingly enjoy the convenience of the hands-free experience. 
KWS and SV are two necessary key technologies of the human-machine interaction system. The machine can open the human-machine dialogue through keyword spotting (KWS) and get authorized dialogue targets based on speaker verification (SV).

KWS aims at detecting predefined keywords in an audio stream~\cite{chen2014small}. In recent years, end-to-end deep neural networks (DNN) have been employed in KWS and achieved superior performance \cite{chen2014small}. Since then, more elaborately designed neural networks are employed to build better performing KWS systems, including convolutional neural networks~\cite{sainath2015convolutional,tang2018deep,choi2019temporal,kim2019temporal}, recurrent neural networks~\cite{fernandez2007application,woellmer2013keyword}, and neural networks based on attention mechanisms~\cite{bai2019time,shan2018attention}, etc.
To improve the detection rate of non-target keywords while maintaining the accuracy of target keywords, \cite{huh2020metric,sacchi2019open} explored the use of deep metric learning methods for the KWS task, where the DNN model is not used directly for classification but rather as a feature extractor that provides specific embeddings of keywords.

SV aims to verify the claimed identity of a person for a given speech~\cite{bimbot2004tutorial}.
In this paper, we only focus on text-independent SV that does not need any restriction on lexical content for speaker modeling as well as testing.
DNNs are widely used for speaker verification because of their ability to extract speaker features effectively ~\cite{snyder2018x,jung2019rawnet}. In particular, for text-independent speaker verification tasks, the speaker features learned by DNNs are independent of the text content.

\begin{figure*}[t]
\vspace{-3em}
  \centering
  \includegraphics[width=\linewidth]{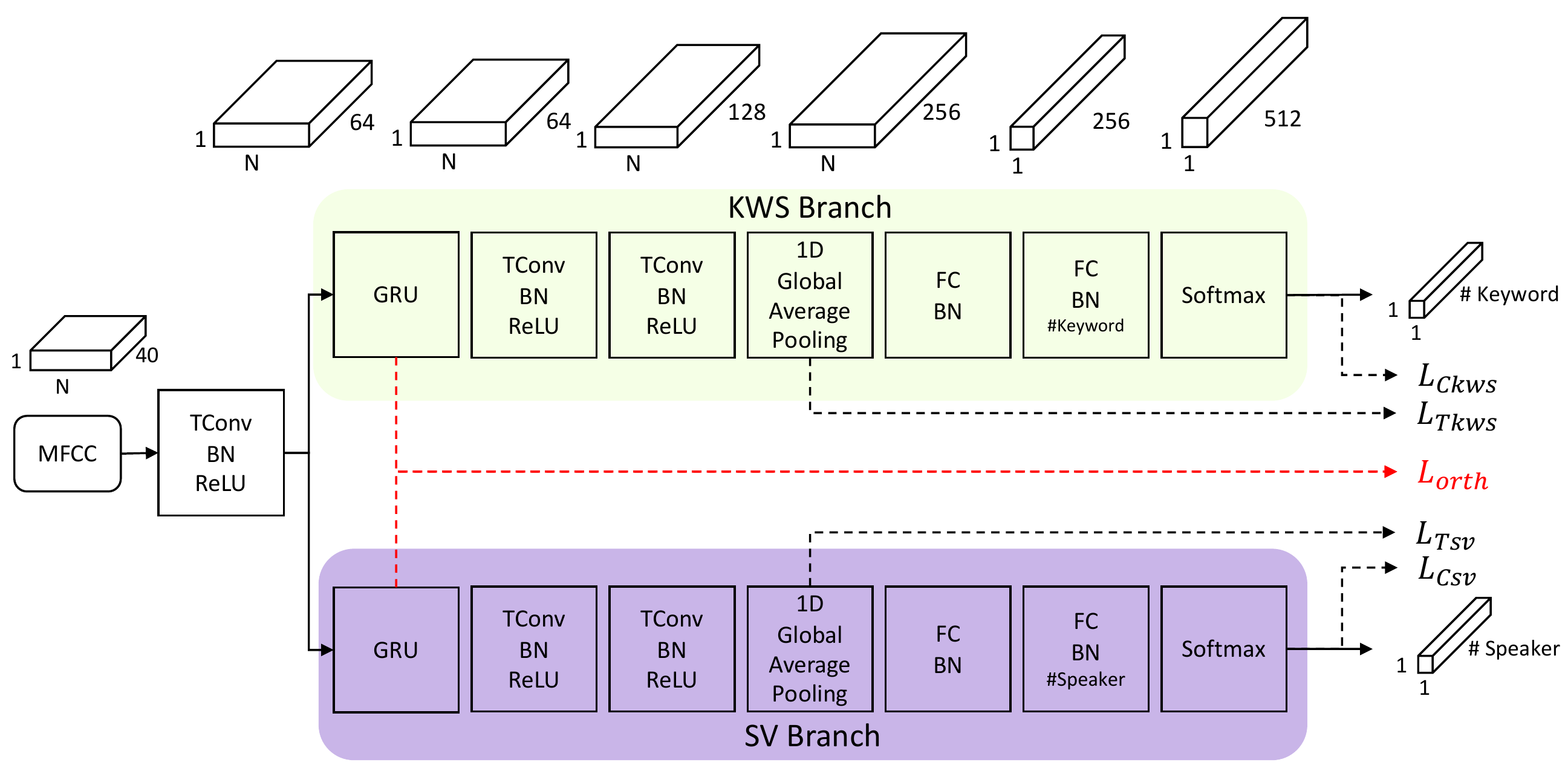}
  \caption{Architecture of the proposed two-branch neural network. It consists of a shared temporal convolutional layer and two branches, the KWS branch and the SV branch. In order to adopt orthogonal regularization, the GRUs of KWS and SV branches were designed with the same neural network structure.
  $N$ is the number of frames before and after layers. The details of $L_{orth}$ is described in Section 2.2.}
  \label{fig:network}
\end{figure*}

Traditionally, the state-of-art KWS and SV models are trained independently using different datasets since they expect to learn distinctive acoustic features. KWS requires features requiring linguistic content as much as possible, while SV requires features with rich speaker information.
However, this independent treatment is not the way how humans process speech signals: humans always simultaneously decipher speech content and paralinguistic information including languages, speaker characteristics and emotions, etc \cite{tang2016collaborative}.
Specifically, for KWS and SV tasks, this ``multi-task processing" relies on three premises:
(1) many common audio features and techniques have been designed and employed for the two tasks, such as Mel-frequency cepstral coefficient (MFCC) features \cite{chen2009speaker,choi2019temporal}, TDNN \cite{snyder2018x,kim2019temporal} modeling framework, and metric learning method \cite{huh2020metric,zhang2018text}; 
(2) all these tasks are basically discriminative tasks, so they can share the same front-end signal processing (e.g. filtering) modules \cite{saremi2016comparative,kumar2018convolutional} and pipeline;
and (3) they are mutual beneficial \cite{tang2016collaborative,liu2020speaker,jung2020multi}, for example, by paying particular attention to specific voices while understanding information about the content of the language, we can verify the speaker's voice; on the other hand, if we are familiar with a speaker, we tend to recognize his/her voice~\cite{liu2020speaker}.

The third point above has been experimentally demonstrated by researchers that these KWS and SV tasks are collaborative \cite{jung2020multi,liu2020speaker}.
Jung \textit{et al.}\cite{jung2020multi} argued that acoustic and speaker domains are complementary. They proposed a multi-task network and introduced global query attention to use the interrelated domain information of KWS and SV. However, the interaction of information between the KWS branch and the SV branch is unidirectional. Speaker features are not used by the KWS branch, which constrains further improvements in its performance.
\cite{liu2020speaker} designed a two-branch neural network, the input is a spectrogram, and the two branches output keyword and speaker results respectively. In order to avoid the interaction between keyword information and speaker information, \cite{liu2020speaker} proposed dual attention to remove the information of another branch from the current branch.

We believe that the key to solve the KWS and SV tasks through a network is to efficiently extract the common features of both tasks and decouple the task-related features.
In this paper, we propose a two-branch neural network to learn the task-specific features of KWS and SV along with the common feature. Orthogonality regularization \cite{bansal2018can} is employed to decouple the linguistic content information for KWS and the speaker information for SV. Thus, under the supervised learning paradigm, the KWS and SV branches can extract more distinct features from each other. Experiments are conducted on Google Speech Commands Dataset (GSCD). The results demonstrate that our proposed method achieve SOTA EER of 1.31\% and 1.87\% on KWS and SV, respectively.

\section{Proposed Method}

\label{sec:pro}

\subsection{Neural Network Architecture}
Inspired by SUDA \cite{liu2020speaker}, we designed a two-branch neural network for KWS and SV. As shown in Fig.\ref{fig:network}, the Mel-Frequency Cepstral Coefficient (MFCC) features are fed into the temporal convolution (TConv in Fig.\ref{fig:network}) to extract the shared features for KWS and SV.
Then, the hidden representation from the first shared layer is passed to the next two gate recurrent unit (GRU) networks that focus on extracting valid information for each of the two sub-tasks, KWS and SV.

\subsection{Decoupling Feature Learning via Orthogonality Regularization}
In this paper, orthogonality regularization is used to decouple latent features between two sub-tasks, KWS and SV.
It aims to decouple the hidden representation to learn speaker-invariant keyword representations and keyword-invariant speaker representations utilizing orthogonality regularization.

Specifically, we apply orthogonality regularization to the GRU of the KWS branch and the SV branch (figure \ref{fig:network} red dashed line).
For each element $x_t$ in the input sequence of time $t$, GRU layer computes the following function
\begin{equation}
\label{form:gru}
\begin{aligned}
&r_{t}=\sigma\left(W_{i r} x_{t}+b_{i r}+W_{h r} h_{(t-1)}+b_{h r}\right) \\
&z_{t}=\sigma\left(W_{i z} x_{t}+b_{i z}+W_{h z} h_{(t-1)}+b_{h z}\right) \\
&n_{t}=\tanh \left(W_{i n} x_{t}+b_{i n}+r_{t} *\left(W_{h n} h_{(t-1)}+b_{h n}\right)\right) \\
&h_{t}=\left(1-z_{t}\right) * n_{t}+z_{t} * h_{(t-1)}
\end{aligned}
\end{equation}
where $h_t$ is the hidden state at time $t$, and $r_t$, $z_t$, $n_t$ are the reset, update, and new gates, respectively. $\sigma$ is the sigmoid function and $*$ is the Hadamard product. $W_{i r}$, $W_{i z}$, $W_{i n}$, $W_{h r}$, $W_{h z}$, $W_{h n}$ are trainable weight matrices.
The following is an example of the computational procedure for computing orthogonal regularization using $W_{i r}$
\begin{equation}
L_{orth}^{i r}=\sum W_{i r}^{kws} {W_{i r}^{sv}}^{T}
\end{equation}
where $W_{i r}^{kws}$ and $W_{i r}^{sv}$ represent the trainable matrices $W_{i r}$ in KWS branch and SV branch, respectively.
Summing up the orthogonal regularization for all the weight matrices in equation~\ref{form:gru} to get the final orthogonal regularization
\begin{equation}
\label{form:orth}
L_{orth}=L_{orth}^{i r}+L_{orth }^{i z}+L_{orth }^{i n}+L_{orth }^{h r}+L_{orth }^{h z}+L_{orth }^{h n}
\end{equation}

\subsection{Loss Function}
The loss function consists of three components, KWS task loss, SV task loss and orthogonality regularization
\begin{equation}
\label{equ:sumloss}
L=L_{kws}+L_{sv}+L_{orth}
\end{equation}
where $L_{kws}$ is the loss for KWS task, $L_{sv}$ is the SV task loss, and $L_{orth}$ is the orthogonal regularization loss, calculated from formula~\ref{form:orth}.
Both $L_{kws}$ and $L_{sv}$ consist of cross-entropy (CE) loss and triplet loss, which provide supervised information for the training process of the model.
\begin{equation}
L_{kws} = L_{Ckws} + L_{Tkws}
\end{equation}
\begin{equation}
L_{sv} = L_{Csv} + L_{Tsv}
\end{equation}
where $L_{Ckws}$ and $L_{Csv}$ denote the CE losses, make the KWS branch extract linguistic content and makes the SV branch extract speaker information. $L_{Tkws}$ and $L_{Tsv}$ are triplet losses, it is employed to increase the  inter-class distance and reduce intra-class distance.
The triplet loss is applied on the 256-dimensional feature vector at the step after 1D global average pooling as shown in Figure \ref{fig:network}, it is calculated in both the KWS and SV branches, the selection of the triplet samples are introduced in Section 3.2.

\section{Experiments}
\label{sec:exp}

\subsection{Dataset}

The Google Speech Commands Dataset version 2 (GSCDv2\footnote{\label{v2}http://download.tensorflow.org/data/speech\_commands\_v0.02.tar.gz}) \cite{warden2018speech} is used this study, which consists of 105,829 utterances of 35 words. All the utterances are spoken by 2,618 speakers.
Every sample has 1 second duration and contains one word.
We use only 2,277 speakers, excluding those with less than 11 utterances. Then disjoint sets of 1959, 159 and 159 speakers are randomly selected for training, validation, and test set, as Table \ref{tab:dataset} shows. Also, to check the robustness of KWS against unseen words, utterances corresponding to the three words `happy', `marvin', and `sheila' are excluded from the training set. In this way, we get 83,636 training samples, 7,644 validation samples and 7,857 test samples from GSCDv2.

\begin{table}[htbp]
  \centering
  \caption{Information on the real-world datasets}
    \begin{tabular}{lrrrr}
    \toprule
      & \multicolumn{1}{l}{Training} & \multicolumn{1}{l}{Validation} & \multicolumn{1}{l}{Test} & \multicolumn{1}{l}{Total} \\
    \midrule
    Utterances & 60,647 & 6,048 & 6,048 & 72,743 \\
    \midrule
    Speakers & 1,213 & 121 & 121 & 1,455 \\
    \bottomrule
    \end{tabular}%
  \label{tab:dataset_xiaomi}%
\end{table}%

In order to verify the validity of the model in the real world, we additionally use the real-world dataset. The keyword is a four-syllable Mandarin Chinese term (“xiao-ai-tong-xue”). We collected 14,543 positive examples and 58,200 negative examples. The splits of the training set, validation set and test set are shown in Table \ref{tab:dataset_xiaomi}. 

\begin{table}[htbp]
\vspace{-1.0em}
  \centering
  \caption{The number of utterances and speakers in training, validation and test set. Extracted from Google Speech Commands Dataset Version 2.}
    \begin{tabular}{lrrrr}
    \toprule
      & \multicolumn{1}{l}{Training} & \multicolumn{1}{l}{Validation} & \multicolumn{1}{l}{Test} & \multicolumn{1}{l}{Total} \\
    \midrule
    Utterances & 83,636 & 7,644 & 7,857 & 99,137 \\
    \midrule
    Speakers & 1,959 & 159 & 159 & 2,277 \\
    \bottomrule
    \end{tabular}%
  \label{tab:dataset}%
\vspace{-1.0em}
\end{table}%

\subsection{Implementation Details}
Our implementation was done with Pytorch deep learning toolkit.
The raw audio is decomposed into a sequence of frames where the window length is 20 ms and the stride is 10 ms for feature extraction.
We use 40-dimensional MFCC.

\textbf{Training.} All the models are trained with a mini-batch of 256 samples using stochastic gradient descent with weight decay of 0.001 and momentum of 0.9.
The initial learning rate is set to be 0.01 and decayed by a factor of 2 when the validation KWS or SV equal error rate (EER) does not decrease for 3 epochs.
The training is terminated when validation KWS or SV EER does not decrease for 10 epochs.
Compared to an enrollment sample, a test sample falls into one of the following four scenarios: 
(1) same keyword, same speaker; 
(2) same keyword, different speaker
(3) different keyword and same speaker;
(4) different keyword, different speaker.
To training efficiency, for each training sample, we have randomly choose 4 samples from the training set, corresponding to the 4 scenarios mentioned above.


\textbf{Evaluation.} 
KWS and SV are basically discrimination tasks that make a decision given a score between embeddings of enrollment and test utterances \cite{jung2020multi}.
EER is used as the metric to evaluate the models in this paper. All utterances in the test set are used once for enrollment.
The cosine distance is used to measure the score.
We train each model 10 times and report its average performance.

\subsection{Baselines}
We use the same structure as the KWS and SV branches to train the KWS and SV tasks separately as our baselines for the KWS and SV tasks to demonstrate the collaborative between the two tasks.
To compare the performance impact of orthogonality regularization and dual attention \cite{liu2020speaker}, we replicated SUDA \cite{liu2020speaker} and trained it on the GSCDv2 dataset following the training method described in Section 3.2.

\subsection{Experimental Results and Analysis}

\subsubsection{Competing and collaborative information properties of KWS and SV}
As shown in the last row of Table \ref{tab:result}, the orthogonality regularization leads to a significant performance improvement.
In addition, formula \ref{form:orth} can be applied to the LSTM with a simple modification, and we also use orthogonality regularization for the LSTM of the KWS branch and SV branch in SUDA. As show in the "SUDA" row of Table \ref{tab:result}, orthogonality regularization brings a consistent performance improvement for proposed model and SUDA.
Our experimental results are consistent with the statement of \cite{tang2016collaborative} that there is collaborative and competitive information between KWS and SV. Good utilization of the information between the two tasks can lead to improved performance of both tasks. Conversely, without orthogonality regularization, the information competition between them makes both tasks perform worse than if they were trained separately, as shown in Table \ref{tab:result}.

\begin{table}[htbp]
\vspace{-1.0em}
  \centering
  \caption{Performance in EER (\%) for the proposed two-branch neural network and baselines on GSCDv2, with 95\% confidence intervals.}
    \begin{tabular}{l|cc}
    \toprule
    Model & KWS & SV \\
    \midrule
    KWS single-task & $1.86\pm 0.10$  & - \\
    \midrule
    SV single-task & - & $2.27\pm 0.03$  \\
    \midrule
    SUDA & $4.87\pm 0.07$  & $5.95\pm 0.08$  \\
    \quad w/ $L_{orth}$ & $2.48\pm 0.11$ & $3.51\pm 0.06$  \\
    \midrule
    Proposed & $\textbf{1.31}\pm 0.03$ & $ \textbf{1.87}\pm 0.07$  \\
    \quad w/o $L_{orth}$ & $2.08\pm 0.05$ & $2.27 \pm 0.09$  \\
    \bottomrule
    \end{tabular}%
  \label{tab:result}%
  \vspace{-1.0em}
\end{table}%

\subsubsection{Impact of training data}

As mentioned in Section 3.2, one test sample falls into one of four scenarios.
In this section, we only keep scenarios (1) same keyword, same speaker and scenarios (4) different keyword, different speaker.
The experimental results are shown in Table \ref{tab:scenarios}, where there is a performance degradation using the training data from two scenarios compared to the training data from four scenarios.
This demonstrates that adding training data from scenarios (2) and (3) allows the model to extract more discriminative features.

\begin{table}[htbp]
\vspace{-1.0em}
  \centering
  \caption{Performance in EER (\%) for different training samples selection, with 95\% confidence intervals.}
    \begin{tabular}{l|cc}
    \toprule
    Training Data & KWS & SV\\
    \midrule
    2 scenarios & $1.71\pm 0.05$ & $2.13\pm 0.12$ \\
    4 scenarios & $\textbf{1.31}\pm 0.03$ & $\textbf{1.87}\pm 0.07$  \\
    \bottomrule
    \end{tabular}%
  \label{tab:scenarios}%
  \vspace{-1.0em}
\end{table}

\subsubsection{Performance in the real world}
We test on real-world dataset, as shown in Table \ref{tab:real-wordresult}, where orthogonality regularization leads to performance improvements.
A noteworthy point is that the difference between the EER of KWS and the EER of SV is about 10 times, which is an issue worth exploring and we will explore this issue in our future work.
\begin{table}[htbp]
\vspace{-1.0em}
  \centering
  \caption{Performance in EER (\%) for the proposed two-branch neural network on real-world dataset, with 95\% confidence intervals.}
    \begin{tabular}{l|cc}
    \toprule
    Model & KWS & SV \\
    \midrule
    Proposed & $\textbf{0.61}\pm 0.04$  & $ \textbf{7.01}\pm 0.14$  \\
    \quad w/o $L_{orth}$ & $0.80\pm 0.07$  & $7.43\pm0.16$  \\
    \bottomrule
    \end{tabular}%
  \label{tab:real-wordresult}%
  \vspace{-1.0em}
\end{table}%

\section{Conclusion}
The performance of keyword spotting (KWS) and speaker verification (SV) tasks can be boosted by leveraging from each other. In this paper, we explore a method to extract the common feature while decoupling task-specific features.
Specifically, we design a two-branch neural network for KWS and SV, and orthogonality regularization is applied to decouple latent features between KWS and SV.
Our proposed method reaches SOTA.
In future work, we will explore the usability of orthogonality regularization in other tasks, such as speaker verification and emotion classification; intent detection and text sentimental classification \cite{huang2021sentiment}.




\bibliographystyle{IEEEbib}
\bibliography{strings,refs}

\end{document}